\documentstyle[12pt]{article}

\def\ben{\begin{enumerate}}  \def\een{\end{enumerate}}
\def\beq{\begin{equation}}   \def\eeq{\end{equation}}
\def\bea{\begin{eqnarray}}  \def\eea{\end{eqnarray}}
\def\nn{\nonumber}
\def\noi{\noindent}

\def\lsim{\raise0.3ex\hbox{$<$\kern-0.75em\raise-1.1ex\hbox{$\sim$}}}
\def\gsim{\raise0.3ex\hbox{$>$\kern-0.75em\raise-1.1ex\hbox{$\sim$}}}

\begin{document}

\begin{center}
{\large \bf The Absolute Definition of the Phase-Shift} \par
{\large \bf in Potential Scattering}
\par

\vspace{0.5 truecm}

{\bf Khosrow Chadan}\\
{Laboratoire de Physique Th\'eorique}\footnote{Unit\'e Mixte de Recherche
UMR 8627 - CNRS }\\    {Universit\'e de Paris XI, B\^atiment 210, 91405
Orsay Cedex, France}

\vspace{0.3 truecm}

{\bf Reido Kobayashi and Takao Kobayashi} \\
{Department of Mathematics \\ Science University of Tokyo, Noda, Chiba 278,
Japan} \\

\vspace{0.8 truecm}
{\it Dedicated to Professor Shinsho Oryu for his 60th anniversary}
      \end{center}

\vspace{0.5 truecm}
\begin{abstract}
The variable phase approach to potential scattering with regular
spherically symmetric
potentials satisfying (\ref{1e}), and studied by Calogero in his book$^{5}$, is
revisited, and we show directly that it gives the absolute definition of the
phase-shifts, i.e. the one which defines $\delta_{\ell}(k)$ as a
continuous function of
$k$ for all $k \geq 0$, up to infinity, where $\delta_{\ell}(\infty
)=0$ is automatically
satisfied. This removes the usual ambiguity $\pm n \pi$, $n$ integer,
attached to the
definition of the phase-shifts through the partial wave scattering
amplitudes obtained
from the Lippmann-Schwinger integral equation, or via the phase of
the Jost functions.
It is then shown rigorously, and also on several examples, that this
definition of the
phase-shifts is very general, and applies as well to all potentials
which have a
strong repulsive singularity at the origin, for instance those which
behave like
$gr^{-m}$, $g > 0$, $m \geq 2$, etc. We also give an example of
application to the
low-energy behaviour of the $S$-wave scattering amplitude in two
dimensions, which
leads to an interesting result.  \end{abstract}

\vspace{1 truecm}

\noi LPT Orsay 01-06 \par
\noi January 2001
 
\newpage
\pagestyle{plain}
\section{Introduction}
\hspace*{\parindent} In quantum scattering theory with a spherically symmetric
potential $V(r)$, the phase-shift, defined for each partial-wave by
the asymptotic
behaviour of the radial wave function at large distances, has the
usual ambiguity of
$\pm \pi n$, $n$ integer. When the potential is regular, i.e. is $L^1(a, \infty
)$, and satisfies the Bargmann-Jost-Kohn condition$^{1,2,3}$

\beq
\label{1e}
\int_0^{\infty} r |V(r)| dr < \infty \quad ,
\eeq

\noi one can show that the phase-shift $\delta_{\ell}(k)$ at infinite
energy satisfies
the condition

\beq
\label{2e}
{\rm tg} \ \delta_{\ell}(\infty ) = 0 \quad , \qquad \hbox{or} \ \sin
\delta_{\ell}(\infty ) = 0 \quad , \eeq

\noi so that one can make the ``canonical'' choice

   \beq
\label{3e}
\delta_{\ell} (\infty ) = 0 \quad ,
\eeq

\noi and then proceed downwards by continuity for finite values of $k$. \par

The same ambiguity exists, of course, when the phase-shift is defined
through the phase
of the Jost function$^{1-3}$. The problem becomes even more serious
when the potential is
singular and repulsive near $r = 0$, and short-range otherwise~:

   \beq \label{4e}
V(r) \ \mathrel{\mathop \simeq_{r \to \ 0}} \ g\ r^{-m}, \ g > 0, \ m > 2\ ; \
\mathrel{\mathop \simeq_{r \to \ 0}}\ g\exp(\alpha /r), \ g > 0,\
\alpha > 0\ ; \
\hbox{etc.} \eeq

In these cases, one can still define the $S$-matrix, $S_{\ell} = \exp (2 i
\delta_{\ell}(k))$, with a continuous phase-shift $\delta_{\ell}(k)$, 
which is again given by the asymptotic behaviour of the wave function 
for $r \to \infty$. However,
now, one has$^{7-9}$

\beq
\label{5e}
\delta_{\ell}(\infty ) = - \infty \quad ,
\eeq

\noi and this adds to the difficulty of finding a unique phase-shift. \par

  From the above remarks, it would seem therefore satisfactory to find a unique
definition of the phase-shift itself by some formula containing
$V(r)$ and the wave
function, and such that (\ref{3e}) is automatically satisfied when
the potential
satisfies (\ref{1e}). And then, to see whether such a definition can
be extended,
eventually with some modification, to potentials which are singular
at $r = 0$ or at $r
= \infty$ (long range). \par

The answer to the above quest has already been found, although not written down
explicitely in the form (\ref{20e}) we give below. It is given by the
variable phase
method$^{5}$. To simplify the algebra, and see what is essential, let
us take the case
of the $S$-wave, $\ell = 0$. We have the radial Schr\"odinger equation

\beq
\label{6e}
\left \{ \begin{array}{l} \varphi '' (k, r) + k^2 \varphi (k, r) =
V(r) \ \varphi (k, r) \\
\\
r \in [0, \infty ) \quad , \quad \varphi (0) = 0\ , \ \varphi '(0)  = 1
\quad . \end{array}\right . \eeq

\noi If we write now the wave function as

\beq
\label{7e}
\left \{ \begin{array}{l} \varphi = A (k, r) \sin [kr + \delta (k,
r)] \quad , \\
\\
A(k, r) \geq 0 \quad , \quad A(k, 0) \not= 0 \quad , \quad \delta (k,
0) = 0 \quad ,
\end{array}\right .
\eeq

\noi it can be shown that one has formally the differential equation$^{5}$

\beq
\label{8e}
\left \{ \begin{array}{l} \displaystyle{{d \over dr}} \ \delta (k, r) = -
\displaystyle{{1 \over k}} \ V(r) \sin^2 [kr + \delta (k, r)] \quad , \\ \\
\delta (k, 0) = 0  \quad , \end{array}\right .
\eeq

\noi and then

\beq
\label{9e}
A(k, r) = {1 \over k} \ e^{{1 \over 2k} \int_0^r V(t) \sin 2 [kt +
\delta (k, t)]dt}
\quad , \eeq

\noi provided (\ref{8e}) has a solution, and the integral in
(\ref{9e}) is finite. We
shall see this in a moment. In essence, $A$ is the amplitude of the
wave function
$\varphi$, whose oscillations are given by $\sin [kr + \delta (k,
r)]$. The function
$\delta (k, r)$ can be interpreted as the local phase-shift since
$\delta (k, R)$ is
the phase-shift due to the cut potential $V(r) \theta (R - r)$. The
total phase-shift
is defined as

\beq
\label{10e}
\delta (k) = \lim_{r \to \ \infty} \delta (k, r) \quad .
   \eeq

In order to see whether (\ref{8e}) has a unique solution, we can write it as

\beq
\label{11e}
\delta (k, r) = - {1 \over k} \int_0^r V(t) \sin^2 \left [kt + \delta
(k, t) \right ] dt \quad .
   \eeq

\noi One can then try to solve this nonlinear integral equation by
iteration, starting
from the zeroth order approximation

\beq
\label{12e}
\delta^{(0)}(k, r) = 0 \quad .
   \eeq

\noi Numerical calculations show that this process is fast converging for usual
regular potentials used in Nuclear Physics, and indeed has been used in
practice. \par

  From (\ref{11e}), it is obvious that the phase-shift is given by

\beq \label{13e}
\delta (k) = - {1 \over k} \int_0^{\infty} V(r) \sin^2 \left [ kr +
\delta (k, r)
\right ] dr \quad .
   \eeq

\noi This formula shows that, for short-range potentials, the tail of the
integral~:
 
\beq
\label{14e}
\left | - {1 \over k} \int_R^{\infty} V(r) \sin^2 \left [ kr + \delta
(k, r) \right ]
dr \right | \leq {1 \over k} \int_R^{\infty} |V(r)|dr \eeq

\noi can be made very small if we choose $R$ larger than the ``range'' of the
potential. \par

In order to study the nonlinear integral equation (\ref{11e}) in a
rigorous way, we can
use the inequality$^{1,2}$

\beq
\label{15e}
|\sin x| \leq C \ {|x| \over 1 + |x|} \quad , \quad \hbox{$x$ real} \quad ,
   \eeq

\noi where $C$ is an appropriate constant. Noting that $x/(1 + x)$ is
an increasing
function of $x$ for $x$ positive, we have

\beq
\label{16e}
|\sin (x + y)| \leq C \ {|x + y| \over 1 + |x + y|} \leq C \ {|x| +
|y| \over 1 + |x| +
|y|} \quad .
   \eeq

\noi Using this in the integral equation (\ref{11e}), we find

\beq
\label{17e}
|\delta (k, r)| \leq {C^2 \over k} \int_0^r |V(t)| \ \left [ {kt +
|\delta (k, t)|
\over 1 + kt + |\delta (k, t)|} \right ]^2 dt \quad , \quad k > 0 \quad . \eeq

\noi It is then obvious that an upper bound $\Delta (k, r)$ for
$|\delta (k, r)|$ is
obtained from the solution of the integral equation

\beq
\label{18e}
\Delta (k, r) = {C^2 \over k} \int_0^r |V(t)| \left [{kt + \Delta (k,
t) \over 1 + kt +
\Delta (k, t)} \right ]^2 dt \quad .
   \eeq

\noi We can again solve this integral equation by iteration, starting from
$\Delta^{(0)} = 0$. It is obvious here that the solution cannot become infinite at any point on
$[0, \infty )$. Indeed, if $\Delta (k, r)$ becomes infinite at $r = r_1$, then, as $r \uparrow
r_1$, the fraction in (\ref{18e}) becomes one, and since $V$ is supposed to be $L^1$, we get a
contradiction. \par

We show in Appendix A that the equation (\ref{18e}) has always a
unique solution for
all values of $k > 0$, provided $V$ satisfies (\ref{1e}), and therefore that
(\ref{11e}) also has a unique solution. We show also that
$\displaystyle{\lim_{k \to
\ \infty}} \Delta (k, \infty ) = 0$, from which we conclude then
directly that we have
(\ref{3e}). However, it is not really necessary to use the integral
equation. Indeed, as
we shall see in the next section, we can express the right-hand side
of (\ref{11e}) in
terms of the regular solution $\varphi$ given by (\ref{6e}), and its
derivative $\varphi
'$, to obtain the new formulae

\beq
\label{19e}
\delta (k, r) = - k \int_0^r V(t) {\varphi^2 (k, t) \over \varphi
'^{\, 2} (k, t) + k^2
\varphi^2 (k, t)} \ dt \quad , \eeq

\noi and

\beq \label{20e}
\delta (k) = - k \int_0^{\infty} V(r) {\varphi^2 \over \varphi'^{\,
2} + k^2\varphi^2} \
dr \quad .
   \eeq

\noi We shall see in the next section how to generalize these
equations to higher
$\ell$. Likewise, the amplitude $A(k, r)$ can also be written as$^{5}$

   \beq
\label{21e}
A(k, r) = {1 \over k} \ \sqrt{\varphi'^{\, 2} + k^2 \varphi^2} \quad .
\eeq

\noi Now, we know, quite generally, that for potentials satisfying
(\ref{1e}), the
wave function $\varphi (k, r)$ exists for all values of $k$, real or
complex, and all
values of $r \geq 0$ $^{1-4}$. Likewise for $\varphi ' (k, r)$, and we
have, of course

\beq
\label{22e}
\varphi '(k, 0) = 1 \quad , \quad \varphi (k, r) \ \mathrel{\mathop
=_{r \to \ 0}}\  r +
r \ o(1) \quad .
   \eeq
\noi Moreover, we have also that $\varphi$ and $\varphi '$, for any
fixed real $k > 0$,
are real and continuous function of $r$, and bounded as $r \to
\infty$. Also, from the
general theory of differential equations$^{12,13}$, we know that, for any $k$,
$\varphi$ and $\varphi '$ cannot vanish simultaneously at any point
$r = r_0 \geq 0$,
since this would entail that $\varphi \equiv 0$, a contradiction with
$\varphi '(k, 0)
= 1$. It follows that, because of (\ref{1e}), the integrals in (\ref{19e}) and
(\ref{20e}) are absolutely convergent for all real values of $k > 0$,
and define
continuous functions of $k$. Therefore, (\ref{19e}) and (\ref{20e})
define, in a very
nice and simple way the ``local'' and the total phase-shift,
respectively. We shall come
in the third section to the case $k = 0$. \par

Several remarks are now in order. First of all, (\ref{20e}) shows that, for a
potential of a given sign, the phase-shift has the opposite sign, something
well-known$^{1,2}$, and also obvious on (\ref{11e}) and (\ref{13e}).
Secondly, it
is obvious on (\ref{19e}) and (\ref{20e}) that the phase-shift is
independent of the
amplitude of $\varphi$. Multiplying $\varphi$ by a constant factor
$\lambda (k)$
independent of $r$ leaves (\ref{19e}) and (\ref{20e}) invariant. This
is as expected, of
course. Finally, from (\ref{21e}) we find

\beq
\label{23e}
A(k,r) > 0 \quad , \quad A(k, 0) = {1 \over k} \not= 0
   \eeq

\noi which was assumed in (\ref{7e}). Now, for $r \to \infty$, the
asymptotic behaviour
of $\varphi$ and $\varphi '$ are given by$^{1-4}$

\beq \label{24e}
\varphi (k, r) \ \mathrel{\mathop \simeq_{r \to \ \infty}} \ {|F(k)|
\over k} \sin (kr +
\delta ) + o(1) \quad ,
   \eeq

\noi and

\beq
\label{25e}
\varphi '(k, r) \ \mathrel{\mathop \simeq_{r \to \ \infty}} \ |F(k)|
\cos (kr + \delta )
+ o(1) \quad , \eeq

\noi where $F(k)$ is the Jost function, which never vanishes for $k >
0$ $^{1-4}$. Using
now (\ref{24e}) and (\ref{25e}) in (\ref{21e}), we find

\beq
\label{26e}
A(k, \infty ) = {|F(k)| \over k} \quad ,
\eeq

\noi and this, used in (\ref{7e}), leads again to (\ref{24e}), a
consistency check.
\par

We have now to see whether, for regular potentials satisfying (\ref{1e}), the
phase-shift defined by (\ref{20e}) satisfies (\ref{3e})~: $\delta
(\infty ) = 0$. This
is very easy to check. Indeed, for $k$ real and going to infinity, we have
uniformly in $r$ $^{1,3,4}$, for any $r \in [0, R]$, $R < \infty$,

\beq
\label{27e}
\varphi (k, r) \ \mathrel{\mathop =_{k \to \ \infty}} \ {\sin kr
\over k} + {1 \over k}
\ o(1) \quad ,  \eeq

\noi and

    \beq
\label{28e}
\varphi '(k, r) \ \mathrel{\mathop =_{k \to \ \infty}} \ \cos kr + o(1) \quad .
\eeq

\noi Using these in (\ref{20e}), together with (\ref{14e}), leads to

\bea
\label{29e}
&&\delta (k) \ \mathrel{\mathop \simeq_{k \to \ \infty}} \ - k
\int_0^{R} V(r) \
{\sin^2 kr \over k^2} \ dr = - \int_0^{R} V(r) \ {1 - \cos 2 kr \over
2k} \ dr \nn \\
&&= -
\int_0^{R} V(r) \ dr \int_0^r \sin 2kt \ dt = - \int_0^{R} \sin 2kt \ dt
\left ( \int_t^{R} V(r) \ dr \right ) \quad . \eea

\noi Now, in general, $W(t) = \int_t^{\infty} V(r) \ dr$ is $L^1(0, \infty )$.
Indeed

   \bea \label{30e}
&&\int_0^{\infty} |W(t)| dt = \int_0^{\infty} dt \left |
\int_t^{\infty} V(r) \ dr \right
| \leq \int_0^{\infty} dt \int_t^{\infty} |V(r)| dr\nn \\
&& = \int_0^{\infty} |V(r)| dr
\int_0^r dt = \int_0^{\infty} r|V(r)|dr < \infty
   \eea

\noi by virtue of (\ref{1e}). Therefore, in the last integral in
(\ref{29e}) we have a
Fourier sine transform of an $L^1$ function. From the
Riemann-Lebesgue lemma, it
follows that, as $k \to \infty$, it vanishes. Therefore, $\delta (k
\to \infty) = 0$.
This argument, which is quite general, can be made more precise (see
Appendix B). It
follows that the definition of the phase-shift given by (\ref{20e})
in terms of the
well-defined regular solution $\varphi$, (\ref{6e}), is indeed an
absolute definition
of $\delta (k)$ for all $k > 0$ satisfying automatically (\ref{3e}). \par

We shall see later that (\ref{20e}) can be extended to potentials
which are outside
the Bargmann-Jost-Kohn class, especially to those potentials which are strongly
repulsive at $r = 0$. We are going now to give first the derivation
of (\ref{8e}),
(\ref{9e}) and (\ref{21e}). These are found essentially in$^{5}$,
except for the integral forms (\ref{19e}) and (\ref{20e}), which are
the basic equations of our paper. We reproduce the proofs for the
convenience of the reader.

\section{Derivation of (19) and (20)}
\hspace*{\parindent} We follow the usual method given in
Calogero$^{5}$ for deriving the
differential equation (\ref{8e}). Differentiating (\ref{7e}), we find

\beq \label{31e}
\varphi ' = A' \sin (kr + \delta ) + A(k + \delta ') \cos (kr +
\delta ) \quad .
\eeq

\noi We have here two unknown functions $A$ and $\delta$, and only
one equation, the
Schr\"odinger equation (\ref{6e}), at our disposal. We can therefore
impose a relation
between $A$ and $\delta$. We impose

\beq
\label{32e}
A' \sin (kr + \delta ) + A \delta ' \cos (kr + \delta ) = 0 \quad .
   \eeq

\noi It follows that (\ref{31e}) becomes simply
\bea
\label{33e}
\varphi ' = Ak \cos (kr + \delta ) \quad .
\eea

\noi Differentiating now this, and using the Schr\"odinger equation
(\ref{6e}), we
find the new equation

\beq
\label{34e}
A' k \cos (kr + \delta ) - Ak \delta ' \sin (kr + \delta ) = V A \sin
(k r + \delta )
\quad .
\eeq

\noi Combining this with (\ref{32e}), we find the differential phase equation
(\ref{8e}) and the amplitude equation (\ref{9e}), a complete set equivalent
to the Schr\"odinger equation (\ref{6e}). \par

However, we can go back again to $\varphi$ and $\varphi '$, using
(\ref{7e}) and
(\ref{33e}). Adding their squares, we find$^{5}$

\bea
\label{35e}
A^2(k, r) = {1 \over k^2} \left ( \varphi '^{\, 2} + k^2 \varphi^2
\right )  \quad .
\eea

\noi It follows that (\ref{8e}) can be written

\bea
\label{36e}
\delta ' &=& - {1 \over k} \ V \sin^2 (kr + \delta ) = - {1 \over k}
\ V {\varphi^2
\over A^2} \nn \\
&=& - k \ V \ {\varphi^2 \over \varphi '^{\, 2} + k^2 \varphi^2} \quad .\eea

\noi This is the basic equation, giving the ``local'' phase-shift
$\delta (k, r)$ in
terms of $\varphi$ and $\varphi '$. Integrating it, we find
 
\beq
\label{37e}
\delta (k, r) = - k \int_0^r V(t) \ {\varphi^2 (k, t) \over \varphi
'^{\, 2} (k, t) + k^2
\varphi^2 (k, t)} \ dt \quad , \eeq

\noi where the initial condition $\delta (k, 0) = 0$ has been used.
Making now $r \to
\infty$, we get the total phase-shift, (\ref{20e}). We have already
checked that
in (\ref{19e}) and (\ref{20e}) the integrals are absolutely
convergent, and define
continuous functions of $k$ for all $k > 0$. \\

\noi {\bf Higher waves.} \\

The method for $\ell \not= 0$ is quite similar$^{5}$. We have to
deal now with the radial Schr\"odinger equation

$$\varphi ''_{\ell} (k, r) + k^2 \varphi_{\ell}(k, r) = \left [ {\ell
(\ell + 1) \over
r^2} + V(r) \right ] \varphi_{\ell} (k, r) \quad , \eqno(38.{\rm a})$$

\noi and its free counterpart when $V = 0$~:

$${d^2 \over dr^2} {u_{\ell} \choose v_{\ell}} + k^2 {u_{\ell}
\choose v_{\ell}} = {\ell
(\ell + 1) \over r^2} {u_{\ell} \choose v_{\ell}} \quad . \eqno(38.{\rm b})$$

\noi These free solutions are given by$^{1-6}$

$$\left \{ \begin{array}{l}
u_{\ell}(kr) = \sqrt{\displaystyle{{\pi kr \over 2}}} \ J_{\ell + {1
\over 2}} (kr) \ \displaystyle{\mathrel{\mathop =_{r \to \ 0}}} \
\displaystyle{{(kr)^{\ell + 1} \over (2 \ell + 1)!!}} + \cdots \quad , \\
\\
v_{\ell}(kr) = \sqrt{\displaystyle{{\pi kr \over 2}}} \ N_{\ell + {1
\over 2}} (kr) \ \displaystyle{\mathrel{\mathop =_{r \to \ 0}}} \
\displaystyle{{(2 \ell
- 1)!! \over (kr)^{\ell}}} + \cdots \quad , \\
\\
(2\ell + 1)!! = \displaystyle{{\Gamma (2 \ell + 2) \over 2^{\ell} \Gamma (\ell + 1)}} \quad ,
\qquad \ell > - {1 \over 2} \quad . \end{array}\right . \eqno(39.{\rm a})$$

\noi We have also

$$\left \{ \begin{array}{l}
u_{\ell}(kr) \ \displaystyle{\mathrel{\mathop =_{r \to \ \infty}}} \ \sin (kr -
\displaystyle{{1 \over 2}} \ell \pi ) + \cdots \\
\\
v_{\ell}(kr) \ \displaystyle{\mathrel{\mathop =_{r \to \ \infty}}} \ \cos (kr -
\displaystyle{{1 \over 2}} \ell \pi ) + \cdots \quad .\\
        \end{array}\right . \eqno(39.{\rm b})$$

\noi Their Wronkian is given by

$$W(u_{\ell} , v_{\ell}) \equiv u'_{\ell} \ v_{\ell} - u_{\ell} \
v'_{\ell} = k \quad .
\eqno(40)$$

\noi  As for the regular solution $\varphi_{\ell}(k,r)$, it is
customary to normalize it
such that$^{1-4}$

$$\varphi_{\ell}(k,r) \ \mathrel{\mathop \simeq_{r \to \ 0}}\
{r^{\ell + 1} \over (2
\ell + 1)!!} + \cdots \eqno(41)$$

Now, again, under the condition (\ref{1e}) on the potential, one can
show that both
$\varphi$ and $\varphi '$ exist for all real or complex values of
$k$, all $r \geq 0$,
and are continuous in both variables$^{1-4}$. Moreover, one has the asymptotic
behaviours

$$\left \{ \begin{array}{l}
\varphi_{\ell} (k, r) \ \displaystyle{\mathrel{\mathop =_{r \to \ \infty}}} \
\displaystyle{{|F_{\ell}(k)| \over k^{\ell + 1}}} \ \sin \left ( kr -
\displaystyle{{1
\over 2}} \ell \pi + \delta_{\ell}(k) \right ) + o(1) \quad , \\
\\
\varphi '_{\ell} (k, r) \ \displaystyle{\mathrel{\mathop =_{r \to \ \infty}}} \
\displaystyle{{|F_{\ell}(k)| \over k^{\ell}}} \ \cos \left ( kr -
\displaystyle{{1
\over 2}} \ell \pi + \delta_{\ell}(k) \right ) + o(1) \quad , \\
   \end{array}\right .
\eqno(42)$$

\noi where $F_{\ell}(k)$ is the Jost function, well-defined and
continuous for all $k
\geq 0$. For $k$ real, one has also, uniformly in $r$ for any $r \in
[0, R]$, $R <
\infty$,

$$\left \{ \begin{array}{l}
\varphi_{\ell} (k, r) \ \displaystyle{\mathrel{\mathop =_{k \to \ \infty}}} \
\displaystyle{{\sin kr \over k^{\ell + 1}}} + \displaystyle{{o(1)
\over k^{\ell + 1}}}
\quad , \\
\\
\varphi '_{\ell} (k, r) \ \displaystyle{\mathrel{\mathop =_{k \to \ \infty}}} \
\displaystyle{{\cos kr \over k^{\ell}}} + \displaystyle{{o(1) \over k^{\ell}}}
\quad . \\
\end{array}\right .
\eqno(43)$$

We write now

$$
    \varphi_{\ell}(k,r) = A_{\ell}(k,r) [u_{\ell}(kr) \cos \delta_{\ell}(k,r) +
    v_{\ell}(kr) \sin \delta_{\ell}(k,r)],
\eqno(44)$$

\noi $A_{\ell}(k,r) \ne  0$ for all $r \ge 0$, $\delta_{\ell}(k,0) =
0$. We have now
again two unknown functions $A_{\ell}  $ and $\delta_{\ell} $, and only one
(differential) equation to determine them. We can therefore impose a
relation between
$A_{\ell}$ and $\delta_{\ell}$. Simplifying the writing, we impose

$$   A '[u \cos \delta  + v \sin \delta  ] + A [ - u \sin \delta  + v
\cos \delta  ]
\delta ' = 0 \eqno(45)$$

\noi Differentiating now (44) and taking into account (45), we find

$$\varphi ' = A [ u ' \cos \delta   + v ' \sin \delta  ] \quad .
\eqno(46)$$

\noi One more differentiation gives us now

$$\varphi '' = A ' [ u ' \cos \delta  + v ' \sin \delta ]  + A [ u
      '' \cos \delta  + v ''  \sin \delta ]$$
$$ + A [ - u ' \sin \delta  + v ' \cos \delta ] \delta ' \quad . \eqno(47)$$

\noi Using now (38.a) for $\varphi ''$, and (38.b) for $u ''$ and $v ''$, we
find from (47)

$$A ' [u ' \cos \delta + v ' \sin \delta  ] + A [- u '\sin \delta + v '
\cos \delta ] \delta '$$
$$= V A [ u \cos \delta  + v \sin \delta  ] \quad  . \eqno(48)$$

\noi We have now two equations, namely (45) and (48) to determine $A$ and
$\delta$. We can write them, symbolically, as

$$a A ' + b A \delta ' = 0 \eqno(49)$$

$$c A ' + d A \delta ' = a V A \quad  . \eqno(50)$$

\noi Eliminating $A '$, and using $A \ne 0$ and (40), we find

$$ad - bc = - (u 'v - u v ') = - k \quad , \eqno(51)$$

\noi and

$$\delta_{\ell} ' = - \frac{1}{k} V(r) [ u_{\ell}\cos \delta_{\ell} +
v_{\ell}\sin
\delta_{\ell}]^2 \quad , \eqno(52)$$

\noi to which we have to add $\delta_{\ell}(k,0) = 0$.
Likewise, eliminating $\delta ' $, we find

$$k A_{\ell} ' = V(r) A_{\ell} [u_{\ell}\cos  \delta_{\ell} +
v_{\ell}\sin \delta_{\ell}
][- u_{\ell}\sin \delta_{\ell} + v_{\ell}\cos \delta_{\ell}  ] \quad .
\eqno(53)$$

\noi As it is easily seen, we have again equations very similar to the
equations for $\ell = 0$. Once (52) is solved, with the boundary condition
$\delta_{\ell}(k,0) = 0$, we can replace its solution $\delta_{\ell}$
in (53), and
integrate it to get $A_{\ell}$~:

$$A_{\ell}(k,r) = A_{\ell}(k,0) \exp \left [ \frac{1}{k} \int_{0}^{r}
\cdots  \right ]
\quad . \eqno(54)$$

\noi However, as we saw previously, we can write both $\delta_{\ell}
$ and $A_{\ell}$
in terms of $\varphi_{\ell}$. From (44) and (46), we can calculate $\sin
\delta_{\ell}$ and $\cos \delta_{\ell}$~:

$$\sin \delta   = \frac{u ' \varphi - u \varphi '}{k A} \eqno(55)$$

$$\cos \delta   = \frac{v \varphi ' - v ' \varphi }{k A}  \quad . \eqno(56)$$

\noi Using now  $\sin ^2 \delta + \cos ^2 \delta  = 1$, we get

$$A_{\ell}^2  =  \frac{1}{k^2} \left [ \left ( u_{\ell} \varphi
_{\ell} ' - u_{\ell} '
\varphi_{\ell } \right )^2 + \left ( v_{\ell} \varphi_{\ell} ' - v_{\ell} '
\varphi_{\ell} \right )^2 \right ] \quad . \eqno(57)$$

\noi Using also (55) and (56) in (52) and remembering (40), we get

$$\delta_{\ell}' = - {1 \over k} \  V(r) \frac{\varphi _{\ell}^2
(k,r)}{A_{\ell}^2(k,r)} \quad .
\eqno(58)$$

Integrating this now, we have

$$\delta_{\ell}(k, r) = - k \int_{0}^{r} V(t) \frac{\varphi _{\ell}^2 (k,t)}{[
     (u_{\ell} \varphi _{\ell} ' - u_{\ell} ' \varphi _{\ell})^2 + ( v_{\ell}
\varphi_{\ell} ' - v_{\ell} ' \varphi _{\ell})^2 ]} \ dt \quad . \eqno(59)$$

\noi All these formulae, so far purely formal, are very similar to those
for $\ell = 0$. However, using now the behaviours of $\varphi$,
$\varphi '$, $u$, $u'$,
$v$, $v'$ for $r \to 0$, we can check that all our above formulae are
meaningful, i.e. the integrals are convergent at $r = 0$. Again, it
is clear from (57)
that $A_{\ell}(k,r) \ne 0$ for all $r \ge 0$. Indeed, if $A_{\ell} $
is zero at $r =
r_{0} $, we must have, according to (57), $u \varphi ' - u ' \varphi
= 0$ and $v \varphi
' - v '\varphi = 0$ at this point. Calculating again $\varphi$  and
$\varphi '$ from
these two equations, we find that, if $k \ne 0, \varphi = \varphi ' =
0$ at $r =
r_{0}$. And this entails that $\varphi \equiv 0$ everywhere, a
contradiction with
(41). Therefore, our assumption on $A_{\ell}$ is satisfied~:
$A_{\ell}(k,r),$ for $ k >
0$,  never vanishes for $r \in [0, \infty)$. \par

      Let us now look at the behaviour of $\delta_{\ell}(k,r)$, (59), $k >
0$ for $ r \to 0$, $\ell > - 1/2$. From (39.a) and (41), it is easily
seen that the numerator behaves like $r^{2\ell + 2}$ whereas the
denominator becomes a positive constant. Therefore, we have, for $k > 0$,

$$\delta_{\ell}(k,r) \ \mathrel{\mathop \simeq_{r \to \ 0}} \
constant \int_{0}^{r} t^{2\ell + 2}
V(t) dt = r^{2\ell + 1} o(1) \quad . \eqno(60)$$

\noi This shows that our assumption on $ \delta_{\ell}(k,r)$ is also
satisfied~: $\delta_{\ell}(k,0) = 0$.  Using (60) in (53), we find
also that we have a convergent integral in (54). This completes the validity of the method and
its consistency. \par

      We have now to look at (57) and (59) for $ r \to \infty$. From
(39.b) and (42), we find easily, for all $k > 0$, that we have

$$A_{\ell}^2 (k,\infty) = \frac{\vert F_{\ell}(k) \vert ^2 }{k^{2\ell + 2}}
    \Rightarrow A_{\ell} (k,\infty) = \frac{\vert F_{\ell}(k) \vert
}{k^{\ell + 1}} \quad
, \eqno(61)$$

\noi where $F_{\ell}(k) $ is the Jost function, finite and continuous
for all $k
\ge 0$ $^{1-4}$. From (59) for $r \to
\infty$, we find

$$\delta_{\ell}(k) = - k \int_{0}^{\infty} V(r) \frac{\varphi_{\ell}^2 (k,r)}{[
     (u_{\ell}' \varphi_{\ell} - u_{\ell} \varphi_{\ell}')^2
+ ( v_{\ell}' \varphi_{\ell} - v_{\ell} \varphi_{\ell}')^2 ]} \ dr
\quad . \eqno(62)$$

\noi and it is easily checked from (39.b) and (42) that the integral here
is well-defined and absolutely convergent, and since $\varphi$,
$\varphi '$, $u$,
$u'$, $v$ and $v '$ are all continuous functions of $k$ for all $k >
0$ $^{1-6}$, the same is true for $ \delta_{\ell}(k)$~: the phase-shift is a
continuous function of $k$ for all $k > 0$. To check (3), we can use
(43) in (62), and
we find, as for the case $\ell = 0$, that $\delta_{\ell}(\infty ) =
0$. Moreover,
$k^{\ell + 1} A_{\ell}$ and $\delta_{\ell}$ are also continuous
functions of $k$ for all
$k > 0$. They are given, respectively, by (57) and (59). The physical
phase-shift is
given by (62), an absolutely convergent integral for all $k > 0$
under the assumption
(1), and one has also (3). Therefore, (62) gives an absolute
definition of the phase-shift $\delta_{\ell}(k)$ in general for all
$k$, and for potentials satisfying (\ref{1e}). We shall see in the
next section that
(\ref{20e}) and (62) are valid also for singular repulsive potentials
as well. The case
$k = 0$ will be considered at the end of the next section. \\

\noindent {\bf Remark.} Making $\ell = 0$, one finds, as expected, all the formulae we found
previously for the $S$-wave. Before ending this
section, let us mention that by combining (38.a) and (38.b), and
integrating from $0$ to $r$, using the appropriate boundary
conditions at $r = 0$, namely (39.a), (40), and (41), to evaluate the integrated terms, one can
calculate $u ' \varphi - u \varphi '$ and $v '\varphi - v \varphi '$,
so that (62) can be written, for $\ell > - {1 \over 2}$, also as

$$\delta_{\ell}(k) = - k \int_{0}^{\infty}
      V(r)\frac{\varphi_{\ell}^2 (k,r)}{(\int_{0}^{r} u_{\ell} \ 
\varphi_{\ell} V
      dt)^2
+ (k^{-\ell} + \int_{0}^{r} v_{\ell} \ \varphi_{\ell} V dt)^2}\  dr  \quad .
\eqno(63)$$

\noi Note here that we have now well precised boundary conditions at $r = 0$ for
all the functions entering in (63), so that care must be taken in using it, whereas in (20) and
(62), the normalization of $\varphi$ at $r = 0$ does not matter. For $-{1 \over 2} \leq \ell <
0$, as we shall see in the next section, (62) and (63) are valid provided we use there the
``distinguished'' pure Bessel solution for $\varphi$, given by the integral equation (68).
Indeed, now, both free solutions $u_{\ell} = \sqrt{r} J_{\ell + {1 \over 2}}(kr)$ and
$v_{\ell} = \sqrt{r} N_{\ell + {1 \over 2}}(kr)$ vanish at $r = 0$, and so, as free solution, we
can start from any combination $a u_{\ell} + b v_{\ell}$, and use it as the inhomogeneous term in
(68). We get then always a solution $\varphi_{\ell}$ with $\varphi(k, 0) = 0$. The
``distinguished'' solution is the one with $b = 0$.

\section{Domain of validity of (20) and (62)}
\hspace*{\parindent}

      As we have seen, formulae (20) for the $S$-wave, or its
generalization (62) for higher waves, are valid for all $k> 0$ and
all $\ell \ge 0$, provided the potential satisfies the integrability
condition (1)~: $r V(r) \in L^{1}(0, \infty)$. Roughly speaking,
this means that $V$ is less singular than $r^{- 2}$ at the origin. We
may now ask whether they are also valid for potentials having stronger
singularity there, for instance $V(r) \sim g r^{-m}$, $m \ge 2$,
   $g> 0$, or $g \exp(\frac{\alpha}{r^{n}})$, $g > 0$, $\alpha >
0$, $n > 0$, etc., as $r \to 0$. Rather than developing
the general formalism for such general singular potentials (singular
and repulsive at the origin), we shall consider explicit examples, and
leave the full theory for a forthcoming paper. \\

{\bf i)} We consider the formula (20) for the $S$-wave, and take boldly
the potential to be the centrifugal barrier

$$V(r) = \frac{\ell(\ell + 1)}{r^2}, \quad \ell > 0 , \eqno(64)$$

\noi in the Schr\"{o}dinger equation (6). The wave function is now just, up
to an unimportant constant multicative factor, $\varphi =
\sqrt{kr}J_{\ell + 1/2}(kr)$ which is of the form $\phi (kr)$. It is
easily seen that
(20) is well-defined because the integral is convergent (absolutely)
both at $r = 0$ and
$r = \infty$. If we make the change of variable $z = kr$ in it, we
find, by writing
$\varphi ' = k \frac{d}{dz} \phi = k \dot{\phi} (z)$, $\dot{} = d/dz$,

$$\delta = - \ell(\ell + 1) \int_{0}^{\infty}\frac{\phi^2 (
    z)}{[\dot{\phi}^2 (z) + \phi^2 (z)]}\ dz \quad ,
\eqno(65)$$

\noi where $\phi = \sqrt{z} J_{\ell + 1/2}(z)$. This last formula is
now independent of
$k$. The same is therefore true for the original formula (20) with
our $\varphi (k,r)$.
We can therefore calculate it at any value of $k$, for instance for $k =
0$. Using (39.a), we find

$$\delta = - \ell (\ell  + 1) \int_{0}^{\infty}\frac{dz}{[(\ell +
1)^2 + z^2]}= - \ell
\int_{0}^{\infty}\frac{dt}{1 + t^2} = - \ell \ \frac{\pi}{2} \quad .
\eqno(66)$$

\noi This is exactly the phase-shift of the centrifugal barrier $\ell (\ell +
1)/r^{2}$ since, without this potential, the wave-function is $\sin
kr$, and with the potential, $\varphi \cong constant \times \sin (kr -
\frac{1}{2} \ell \pi)$, as $ r \to \infty$, according to (42).  \par

In conclusion, our formula (20) is valid for repulsive singular
potentials $g/r^{2}$, $g > 0$, which violate (1) both at $r = 0$
and $r = \infty$.\\

{\bf ii)}  We consider now the previous example, but with $-1/2 \leq \ell <
0$. Here, proceeding as before, we find, as expected, again (66).
Note that $\delta$
is now po\-si\-ti\-ve because the potential  $\ell ( \ell + 1)/r^2$
is negative. For
$\ell = - {1 \over 2}$, the full solution is $\varphi = \sqrt{kr}
J_0(kr)$, so that $\varphi
\sim \sqrt{r}$ and $\varphi ' \sim 1/(2 \sqrt{r})$ as $r \to 0$. \\

\noi {\bf Remark.} Formula (20) was proved for regular potentials. In
all rigor, in
order to apply it to the centrifugal barrier potential $\ell (\ell +
1)/r^2$, we must
first regularize this at the origin, for instance by cutting it by $\theta (r -
\varepsilon )$, and then make $\varepsilon \downarrow 0$, or by
replacing $r$ in the
denominator by $(r + \varepsilon )$, and again take the limit
$\varepsilon \downarrow
0$. However, as we saw, at the limit, we have already an absolutely
convergent integral
for all $\ell \geq - {1 \over 2}$. Also, the derivation of (20) was
based on the
assumption $\varphi '(0) = 1$, i.e. $\varphi (r) \simeq r + \cdots$
for $r \to 0$.
However, for $\ell \not= 0$, we have rather $\varphi_{\ell} \simeq
r^{\ell + 1} +
\cdots$. The extra factor $r^{\ell}$ comes from the regularized formula for the
amplitude $A$ in the limit $\varepsilon \downarrow 0$, as can easily
be seen$^{5}$. \\

{\bf iii)}  We consider now (13) with

$$V(r) = \frac{\ell(\ell + 1)}{r^{2}} + V_{1}(r)\ , \quad  \ell \ge -
\frac{1}{2}\quad ,
\eqno(67)$$

\noi assuming that $r V_{1}(r) \in L^{1}(0 , \infty)$. Consider first $\ell > - {1 \over 2}$. As
we saw, the presence of this
``weak perturbation'', as compared to $\ell(\ell + 1)/ r^{2}$, does
not modify the
behaviour of the regular solution $\varphi_{\ell}$ at $r = 0$, given
by (41), to be
compared with (39.a)$^{1-4}$. And since (20), as we just saw, works
with $\ell(\ell +
1)/r^{2}$, $\ell \ge - 1/2$, applied to the regular solution
$\varphi_{\ell}$, it should
work also  when we add $V_{1}$, provided we use always pure Bessel
functions as free
solutions. This means that the Volterra integral equation which combines the
Schr\"odinger equation and the boundary condition at $r = 0$ is

$$\left \{ \begin{array}{l} \varphi_{\ell}(k, r) = {1 \over k^{\ell + 1}} u_{\ell}(kr) + \int_0^r
G_{\ell}(k;r,r') \ V(r') \ \varphi_{\ell} (k, r') \ dr' \quad , \\
\\
G_{\ell}(k;r,r') = {1 \over k} \left [u_{\ell} (kr) v_{\ell} (kr')
- u_{\ell}(kr')
v_{\ell}(kr) \right ] \quad . \end{array} \right . \eqno(68)$$

We consider now $\ell = - {1 \over 2}$. As has been shown in reference$^{10}$, in order to
formulate a decent scattering theory leading to the asymptotic form (42) with a well-defined
Jost function $F_{\ell}(k)$ and a well-defined phase-shift $\delta_{\ell}$, one has to make
stronger assumptions on $V$ than (1), namely

$$\left \{ \begin{array}{l} \displaystyle{\int_0^{\infty}} r|V(r)| (1 + |\log r|) dr < \infty
\quad , \\ \\
\displaystyle{\int_a^{\infty}} r |V(r)|(\log r)^2 dr < \infty  \quad , \quad a > 0 \quad .
\end{array} \right . \eqno(69)$$

\noi This will be used later for $\ell = - 1/2$, and $k
\to 0$, where we give more details. \par

In conclusion, formula (20) is valid for (67) and the
regular solution $\varphi_{\ell}$. Therefore, we have now two methods to
deal with (67). The first one is to apply (62) to $V_{1}$, and the
second one to apply (20) to the full potential (67). In any case, we
get, of course, the full phase-shift

$$\delta_{\ell}^{total}= \delta_{\ell} - \frac{1}{2} \ \ell \pi \quad ,
\eqno(70)$$

\noi $\delta_{\ell}$ being the physical phase-shift due to $V_{1}$, which is
what interests us in scattering theory. Remember that, in (20) or
(62), $\varphi$ is
always the full solution~: solution of (38.a) with $V = V_1$, or (6)
with (67), always
together with (41).\\

{\bf iv)}  We consider now more singular potentials, namely $V(r) =
g/r^{m}$, $m >2$,
$g > 0$. Here, it is known that
    the solution of the Schr\"{o}dinger equation (1) which vanishes at
    $r = 0$ behaves there as$^{7-9}$

$$\varphi_{\ell} (k, r) \ \mathrel{\mathop \simeq_{r \to \ 0}} \
[V(r)]^{- 1/4} \exp
\left ( - \int_{r}^{\infty}[V(t)]^{1/2}dt \right )$$
$$= \phi_0 (r) = g^{- 1/4}\ r^{m/4} \exp \left [- \sqrt{g}\ {2 \over m - 2}
\ \frac{1}{r^{\frac{m - 2}{2}}} \right ] \quad , \eqno(71)$$

\vskip 5 truemm

\noi independent of $k$ and $\ell$. The wave function $\varphi$ and all of its
derivatives $\varphi^{(n)}$ vanish exponentially at $r = 0$.
Notice that we can omit the factor $g^{- 1/4}$ in front of the last
expression since our formulae for the phase-shifts are homogeneous in
$\varphi$. In fact, at $k = 0$, the Schr\"{o}dinger equation is
soluble exactly, and its solution is$^{7}$, up to an unimportant constant multiplicative factor,

$$\varphi_{\ell}(k = 0, r) = \sqrt{r}\ K_{\frac{2\ell + 1}{m -
2}}\left ( \frac{2
    \sqrt{g}}{m - 2}\ r^{- \frac{m - 2}{2}} \right ) \quad ,
\eqno(72)$$

\vskip 5 truemm

\noi where $K_{\nu}$ is the modified Hankel function$^6$. Using the asymptotic
behaviour of $K_{\nu}(x)$ for $x \to \infty$, we find indeed,
up to constant multiplicative factors, the behaviour shown in (71). On
the other hand, we know that now, because of the strong singularity of
$V$ at $r = 0$, the phase-shifts does not go to zero as $k \to
\infty$, contrary to the case of regular potentials satisfying (1).
Rather, one has the high energy behaviour$^{8,9}$

$$\delta_{\ell}(k) \mathrel{\mathop =_{k \to \ \infty}} - A \ g^{\frac{1}{m}}
\ k^{\frac{m-2}{m}} +  \cdots \eqno(73)$$

\noi where

$$A = \frac{\sqrt{\pi}}{2} \ \frac{\Gamma ( 1- 1/m)}{\Gamma (3/2 -
1/m)} \quad .
\eqno(74)$$

\vskip 5 truemm

\noi Since the main term in (73) is independent of $\ell$, we shall
consider the case
$\ell = 0$ to simplify the algebra, and therefore use (20). In this
formula, the
integral can be split into $\int_0^R + \int_R^{\infty}$. For the
second integral, we
have

$$\left | k \int_{R}^{\infty} \cdots \right | < {g \over k}
\int_{R}^{\infty} {dr
\over r^m} = O \left ( k^{{m-2 \over m} + \varepsilon} \right ) \quad
,  \eqno(75)$$

\noi $\varepsilon$ very small ($> 0)$, as $k \to \infty$, provided $R
= O(k^{-{2 \over m}
+ \varepsilon})$. The contribution of (75) can therefore be neglected
if we compare with
(73). Note that $R \to 0$ as $k \to \infty$. In the first integral,
we can therefore
replace, in first approximation, $\varphi$ by $\phi_0$ given in (71),
and independent
of $k$. We find then

$$ - g k \int_0^R {1 \over r^m} \ {\phi_0^2 \over \phi{'}_{0}^{2} +
k^2 \ \phi_0^2} \ dr
= - g k \int_0^R {1 \over \left ( {m \over 4} \ r^{{m \over 2} - 1} +
\sqrt{g} \right
)^2 + k^2 r^m} \ dr \ . \eqno(76)$$

   \noi Making now the change of variable $x = k^{2/m}
r$, letting $k \to \infty$, and noting that $k^{2/m} R \to \infty$, we find

$$- gk \int_0^R \cdots \mathrel{\mathop \simeq_{k \to \ \infty}}
-gk^{{m-2 \over m}}
\int_{0}^{\infty} {dx \over g + x^m} = - g^{1/m} k^{{m-2 \over
m}} \int_{0}^{\infty}{dt \over 1 + t^{m}} \quad . \eqno(77)$$

It follows that the approximate value of $\delta (k)$ behaves asymptotically as

$$
\delta_{app}(k) \mathrel{\mathop \simeq_{k \to \ \infty}} -B \
g^{\frac{1}{m}} \
k^{\frac{m - 2}{m}} \quad , \quad B = \int_0^{\infty} {dt \over 1 + t^m} \quad
.\eqno(78)$$

\noi This coincides with (73) up to a numerical factor, and is
obtained without much
effort, as we see. \par

The above argument to obtain (78) is, of course,
heuristic because (71) is uniform in $k$ only for $k \in [0, K]$, $K
< \infty$. But it
can be made more precise and quite rigorous. In a forth-coming paper,
we shall develop a
technique to deal with all these problems in a unified way.\\

\noi {\bf Remark}. The remark at the end of {\bf ii)}  applies here
too. We must regularize first the potential, and then let
$\varepsilon\downarrow 0$. The exponential decrease of $\varphi$ as
$r\rightarrow 0$ comes then from the amplitude $A$ in the limit
$\varepsilon\downarrow 0$.\\

\newpage
\noi {\bf A low-energy example.} \\

Our last example is the low energy behaviour of the phase-shift
for $\ell = - 1/2$, where the potential is assumed to be repulsive $(V \ge
0)$, and to satisfy the integrability conditions

$$\int_{0}^{\infty} r \vert V(r) \vert (1 + \vert \log r \vert ) dr <
\infty \quad , \eqno(79)$$

\noi and

$$\int_a^{\infty} r \vert V(r) \vert  (\log r)^2   dr < \infty \quad  ,
\quad a > 0 \quad .
\eqno(80)$$

\noi This corresponds to the $S$-wave Schr\"odinger equation in two
space dimensions,
and is interesting to study$^{10}$. \par

As we saw in {\bf iii)}, we must use here only the pure ``Bessel''
solutions. This means that the solution $\varphi$ is the solution of
the integral equation (68) for $\ell = - {1 \over 2}$~:

$$\varphi(k,r) = \sqrt{kr} \ J_{0}(kr) + \int_{0}^{r} G(k,r,r')\ V(r')\
\varphi(k,r') \ dr' \quad ,
\eqno(81)$$

\noi where the Green's function is given by

$$G = \frac{\pi}{2} \sqrt{rr'} \ \left [ J_{0}(kr)\ N_{0}(kr') -
J_{0}(kr')\ N_{0}(kr) \right ] \quad . \eqno(82)$$

\noi The occurrence of $\log r$ in (79) is due to the presence of
$\log (kr)$ in $N_0$. Here, $\varphi$ is normalized somewhat differently, but we know that the
normalization of $\varphi$ does not matter in (20), (62). In (63), $\varphi$ is normalized
now as to $\varphi \simeq \sqrt{r}$ for $r \to 0$.\par

It is then shown in the above reference that the phase-shift, i.e.
the  phase-shift
$\delta_{0}(k)$ of the $S$-wave in the two dimentional space problem,
has the universal
behaviour

$$\delta_{0}(k) \mathrel{\mathop \simeq_{k \to \ 0}} \frac{ -
\frac{\pi}{2}}{\vert
\log  k \vert} + \cdots \quad ,
\eqno(83)$$

\noi i.e. the main term is independent of the potential.\par

   We are going to find (83) by using the formula (63) for $\ell = -
1/2$, in which
$\varphi_{\ell}$ is given as above, and, for $z \to 0$, $^6$

$$\left \{\begin{array}{l}
J_0(z)  = 1 - \displaystyle{{z^2 \over 4}} + \cdots \\ \\
N_0(z) = \displaystyle{{2 \over \pi}} \left [ \log \displaystyle{{z
\over 2}} + \gamma
\right ] J_0 (z) + O(z^2) \quad . \end{array} \right . \eqno(84)$$

\noi From the integral equation for $\varphi$, it can be easily shown
that, under the
assumptions (79) and (80), the low-energy behaviour of $\varphi$ is
given by $\varphi
\displaystyle{\mathrel{\mathop \simeq_{k \to \ 0}}} \sqrt{kr} J_0(kr)
\simeq \sqrt{kr} +
\cdots$, which we have to normalize to $\varphi \simeq \sqrt{r}$, and, for $u$ and $v$, in order
to comply with (40), we must take

   $$\left \{
\begin{array}{l} u \simeq \sqrt{kr} + \cdots \\
v \simeq- \sqrt{kr}\log kr + \cdots \quad . \end{array} \right . \eqno(85)$$

\noi Using now the above low-energy behaviours in (63), we find

$$\delta (k) \ \mathrel{\mathop \simeq_{k \to \ 0}} \ -
\int_{0}^{\infty} {rV(r) \over \left ( \int_{0}^{r}t V(t) dt
\right)^2 + \left (1 -
\int_0^r tV(t) \log kt \ dt \right )^2} \ dr \ . \eqno(86)$$

\noi Since $V$ was assumed to be positive, we can introduce the new
variable $X = X(r)$ by

$$X(r) = |\log k| \int_{0}^{r} t V(t) dt \quad , \quad  dX =
|\log k| r
V dr \quad , \eqno(87)$$

\noi which is a one-to-one mapping from $r \in [0, \infty)$ to $X \in [0, |\log k|A]$,
where $A = \int_0^{\infty} r|V(r)|dr$. Letting now $k \to 0$, we find easily from (86) 

$$\delta (k) \ \mathrel{\mathop \simeq_{k \to \ 0}}\   -
  {1 \over |\log k|} + \cdots \quad . \eqno(88)$$

\noi Here again, as for the case of singular potentials, we do not get the exact constant $\pi
/ 2$ because we use $\varphi \simeq \sqrt{r}$, which is not uniform on the entire $r$-axis. We
shall come back to this in more detail in the forthcoming paper. \\

\noi {\bf The case k = 0}. \\

      Let us consider now  $k = 0$ in (20). It is known that, under
(1), one has$^{1-4}$

$$\varphi (0,r) \mathrel{\mathop =_{r \to \ \infty}} \varphi '(0,
\infty) r + D + o(1)
\quad , \eqno(89)$$

\noi where $\varphi '(0, \infty)$ is finite or zero, and $D$ is finite
also. Making now $k = 0$ in (20), we find, for the $S$-wave scattering
length$^{1,2}$

$$a_{0} = \lim_{k \to \ 0} \frac{ - \delta_{0}(k)}{k} =
   \int_{0}^{\infty} \frac{\varphi ^2 (0,r)}{\varphi^{'2}(0,r)} V(r) dr \quad .
\eqno(90)$$

At the origin $r = 0$, there is no convergence problem since $\varphi
'(k, 0) =  1$
for all $k$. However, at $r = \infty$, because of (89), we must assume

$$\int_{R}^{\infty} r^2 \vert V(r) \vert dr < \infty
\eqno(91)$$

\noi in order to secure proper convergence, which is also
well-known$^{1,2}$. But this
is not yet the end. We must also be sure that $\varphi '(0, r)$ does
not vanish for $r > 0$. This is surely the case if $V$ is positive$^{1,2}$, but cannot
be guaranteed otherwise. In conclusion, (91) is valid only when $\varphi '(0, r)
\not= 0$ for all $r > 0$. For higher waves, in order to have proper convergence at $r = \infty$,
one needs$^{1,2}$

$$\int_0^{\infty} r^{2\ell + 2} \ |V(r)|dr < \infty  \eqno(92)$$

\noindent and again the non-vanishing of the denominators in (20) or
(62). We shall see
in a forthcoming paper, how to modify these formulae in the presence
of bound states.
However, we know that in all cases, $\delta_{\ell}(k)$ is continuous
down to $k + 0$,
and one has the Levinson theorem $\delta (+ 0) = n \pi$, where $n$ is
the number of
bound states$^{1-5}$. \\

\noi {\bf Two-potential case.} \\

Formulae (62) and (63), as it is obvious, can of course be applied in
the case where we
have two potentials~:

$$V = V_1 + V_2 \quad , \eqno(93)$$

\noi both satisfying (1). Here, we can apply either (20) to $V$, or
(62) and (63) to
$V_2$, where $u_{\ell}$ and $v_{\ell}$ are now replaced by two
appropriate independent
solutions of the Schr\"odinger equation $\varphi_1$ and $\psi_1$,
normalized according
to (40)~:

$$W[\varphi_1, \psi_1] = \varphi'_1 \psi_1 - \varphi_1 \psi '_1 = k
\quad . \eqno(94)$$

\noi With (20), we would get the full phase-shift, and with (62) or (63), the
phase-shift $\delta_2$ due to $V_2$.

\newpage
\noindent
{\Large\bf Appendix A} \\

We have to study here the integral equation (18)~:

$$\Delta (k, r) = {D \over k} \int_0^r |V(t)| \left [ {kt + \Delta
(k,t) \over 1 + kt +
\Delta (k, t)} \right ]^2 dt \quad . \eqno({\rm A}.1)$$

\noi We solve it by iteration, starting from

$$\Delta^{(0)}(k, r) = 0 \quad . \eqno({\rm A}.2)$$

\noi Since $x/(1 + x)$ is an increasing function of $x$ for $x \geq
0$, we get the
increasing sequence of iterations

$$\Delta^{(1)}(k, r) < \Delta^{(2)}(k, r) < \cdots \quad , \eqno({\rm A}.3)$$

\noi where

$$\left \{ \begin{array}{l} \Delta^{(1)}(k, r) = \displaystyle{{D \over k}}
\displaystyle{\int_0^r} |V(t)| \left ( \displaystyle{{kt \over 1 +
kt}}\right )^2 dt
\quad , \\
\\
\Delta^{(n)}(k,r) = \displaystyle{{D \over k}}
\displaystyle{\int_0^r} |V(t)| \left ( \displaystyle{{kt +
\Delta^{(n-1)}(k,t)  \over 1 +
kt + \Delta^{(n-1)} (k,t) }}\right )^2 dt \quad . \\
\end{array}\right . \eqno({\rm
A}.4)$$

\vskip 5 truemm
 
{\bf i)} Assume now first that $V$ is integrable at $r = 0$ and therefore is
$L^1(0, \infty )$.
It is then obvious that the increasing sequence of iterations is bounded by

$$\overline{\Delta}(k,r) =  {D \over k} \int_0^r |V(t)| dt \quad ,
\quad r \geq 0\ , \
k > 0 \quad . \eqno({\rm A}.5)$$

\noi It has therefore a limit, and one has the solution

$$\Delta(k,r) = \lim_{n \to \ \infty} \Delta^{(n)}(k, r) \leq
\overline{\Delta}(k, r)
\eqno({\rm A}.6{\rm a})$$

\noi for all $r \geq 0$, and all $k > 0$. For $r \to \infty$, the
same statement is
valid. Indeed, if we note that both $\Delta (k, r)$ and
$\overline{\Delta} (k, r)$ are
increasing functions of $r$, it follows that we have, when $r \to \infty$,

$$\Delta(k) = \lim_{r \to \ \infty} \Delta(k, r) \leq
\overline{\Delta}(k) = \lim_{r \to
\ \infty} \overline{\Delta}(k, r) = {D \over k} \int_0^{\infty}
|V(t)|dt \ .   \eqno({\rm
A}.6{\rm b})$$

\noi It is then obvious from (A.5) that we have $\Delta (k \to \infty
) = 0$, which gives
in turn (3), as expected. \par

{\bf ii) } If we have only (1)~: $r V(r) \in L^1$, we must refine slightly our
argument. Since our
problem is now the convergence of the integral in $\Delta^{(n)}(k,
r)$ at $r = 0$, we
consider $r$ very small. Now, as it is obvious, another sequence of
upper bounds for
$\Delta^{(n)}(k, r)$ is obtained by using the drastic inequality
$x/(1 + x) < x$ in
(A.1). We obtain in this way a sequence of upper bounds for
$\Delta^{(n)}$ from the
sequence of iterations of

$$\overline{\overline{\Delta}}(k,r) = {D \over k} \int_0^r |V(t)| \left [ kt +
\overline{\overline{\Delta}}(k, t) \right ]^2 dt  \quad .
\eqno({\rm A}.7)$$

\noi Putting $\overline{\overline{\Delta}} = k \omega$, we get

$$\omega (k,r) = D \int_0^r |V(t)| (t + \omega )^2 dt \quad ,
\eqno({\rm A}.8)$$

\noi which is in fact independent of $k$, and must be iterated now.
The above equation
is nothing else than the integral equation for minus the local
scattering length

$$a(r) = \lim_{k \to \ 0} {- \delta (k, r) \over k}
\eqno({\rm A}.9)$$

\noi for the potential $- D|V(r)|$, and has been studied in the book of
Calogero$^5$, chapters 11 and 12, where it is shown that the
iteration of (A.8) leads
to an absolutely convergent series expansion for the solution,
provided $r$ is small
enough and $r V(r) \in L^1$. Alternatively, (A.8) is nothing else
than the Riccati
equation

$$\omega '(r) = D|V(r)| [r + \omega (r) ]^2 \eqno({\rm A}.10)$$

\noi with $\omega (0) = 0$, which has been also thoroughly studied in
the books of
Hille$^{12}$ and Coddington and Levinson$^{13}$, to which we refer
the reader, with the
same conclusion. \par

Once we secure the solution of (A.7) in a small interval $[0, r_0]$, with $\Delta (k, 0) = 0$, we
can then start at $r = r_0$, consider instead of (A.1) the integral equation  

$$\Delta (k, r) = \Delta (k, r_0) + {D \over k} \int_{r_0}^r |V(t)| \left [ {kt + \Delta (k, t)
\over 1 + kt + \Delta (k, t)} \right ]^2 dt  \quad , \eqno({\rm A}.11)$$

\noi and proceed as before, by iteration. Here, we need only $V \in L^1(r_0, \infty )$. We get
then again an increasing sequence bounded by

$$\Delta (k, r_0) + {D \over k} \int_{r_0}^{\infty} |V(t)| dt \quad , \eqno({\rm A}.12)$$

\noi with the same conclusions as before for the existence of the limit of the iterations, ...
etc. However, the high energy behaviour of $\delta (k, r)$ cannot be obtained from the above
analysis. \par

A different method, which superseeds the above considerations, and provides at the same time
the high energy limit of $\delta (k, r)$ and of $\delta (k) = \delta (k, \infty )$ is as
follows. It consists in neglecting $\Delta$ in the denominator of the right-hand side of
(A.1)~:  

$$\Delta (k, r) \leq {D \over k} \int_0^r |V(t)| \left [ {kt + \Delta (k, t) \over 1 + kt}
\right ]^2 dt \leq {D \over k} \int_0^r {r \over t} |V(t)| \left [ {kt + \Delta \over 1 + kt}
\right ]^2 dt \quad . \eqno({\rm A}.13)$$

\noi Writing now $\Delta = r \omega$, the two ends of (A.13) lead us to the integral equation

$$\omega = {D \over k} \int_0^r t |V(t)| {(k + \omega )^2 \over (1 + kt)^2} \ dt \eqno({\rm
A}.14)$$

\noi whose solution provides still a stronger upper bound for $\Delta (k, r)$. Now, it is
trivial to solve (A.14). We just differentiate it, and integrate the differential equation,
taking into account $\omega (k, 0) = 0$. This solution is just~:

$$\left \{ \begin{array}{l} \omega (k, r) = \displaystyle{{k \ I(k, r) \over 1 - I(k, r)}}\quad
,\\ 
\\
I(k, r) = D \displaystyle{\int_0^r} \displaystyle{{t|V(t)| \over (1 + kt)^2}} \ dt  \quad .
\end{array} \right . \eqno({\rm A}.15)$$

\noi The solutions exists as long as $I(k, r) < 1$, that is, in the interval $[0, r_0)$, where
$r_0(k)$ is given by

$$D \int_0^{r_0} {r |V(r)| \over (1 + kr)^2} \ dr = 1 \quad . \eqno({\rm
A}.16)$$

\noi Note that, in any fixed interval $[0, R]$, $R \leq \infty$, we have

$$\lim_{k \to \, \infty} I (k, r) = 0 \quad . \eqno({\rm A}.17)$$

\noi Indeed, from (A.15), and using $1 + kt \geq 1$ and $1 + kt > kt$, we have 

$$I(k,r) \leq I(k, \infty ) \leq D \int_0^{\infty} {t|V(t)| \over 1 + kt} \ dt = J(k, \infty )
\leq \int_0^{\varepsilon} t|V(t)| dt + {1 \over k} D \int_{\varepsilon}^{\infty} |V(t)| dt
\quad , \eqno({\rm A}.18)$$

\noi where, in obvious notations, $J(k,r)$ is defined by $D\int_0^r[t |V(t)|/(1 + kt)]dt$. Now,
we can make first $\varepsilon$ small enough, independent of $k$, in order to make the first
integral in the right-hand side as small as we wish. Once $\varepsilon$ is fixed, we can than
make $k$ large enough in order to make also the second integral as small as we wish. This proves
(A.17). \\

\noi {\bf Remark.} It is obvious that (A.17) is uniform in any finite interval $0 \leq r \leq
R$. \\

As a consequence of (A.17), $r_0(k)$ defined by (A.16) satisfies

$$\lim_{k \to \, \infty} r_0(k) = \infty \quad , \eqno({\rm A}.19)$$ 

\noi so that, the larger $k$ is, the larger is the domain of validity of (A.15). In any case,
by making $k$ large enough, and combining (A.15) and (A.17), we can have, uniformly in $r$ in
$[0, R]$, 

$$\omega (k, r) \leq 2k\ I(k, r) \quad , \qquad k > K \quad , \eqno({\rm A}.20)$$

\noi and we know that $r\omega (k,r)$ is an upper bound for $\Delta (k, r)$. We can therefore,
for $k$ large enough, use (A.20) in (A.1)~:

$$\Delta (k, r) = {D \over k} \int_0^r |V(t)| [ \cdots ]^2 dt \leq {D \over k} \int_0^r |V(t)|
[\cdots ]^1 dt \leq {D \over k} \int_0^r |V(t)|$$
$${kt + 2kt\ I(k, t) \over 1 + kt} \ dt = J(k, r) + 2D \int_0^r {t|V(t)| \over 1 + kt} \ I(k,
t) dt \leq J(k, r) + 2I (k, r) \ J(k, r) \quad . \eqno({\rm A}.21)$$

\noi Now, as we saw for (A.17), both $I$ and $J$ go to zero as $k \to \infty$, uniformly in $r$
for $r$ in any finite interval $[0, R]$. Therefore 

$$\lim_{k \to \, \infty} \Delta (k, r) = 0 \quad , \qquad 0 \leq r \leq R \quad . \eqno({\rm
A}.22)$$

Once we have shown the existence of the solution $\Delta (k, r)$ in $[0, R]$, we can proceed as
for (A.11), write, for $r > R$,

$$\Delta (k, r) = \Delta (k, R) + {D \over k} \int_R^r |V(t)| \left [ {kt + \Delta (k, t) \over 1
+ kt + \Delta (k, t)} \right ]^2 dt \quad , \eqno({\rm A}.23)$$

\noi and proceed again by iteration. This way of proceeding is legitimate since (A.1), or
(18), was obtained from the differential equation (8) and the bound (16). We can therefore
start at any point $r = R$, and integrate after, provided we know $\delta (k, R)$. And we
obtain a bigger upper bound if we replace $|\delta (k, R)|$ by $\Delta (k, R)$. The process is
now very similar to what we had in {\bf i)}. We get an increasing sequence of iterations
$\Delta^{(n)}(k, r)$, with $\Delta^{(0)}(k, r) = \Delta (k, R)$, and a global upper bound for all
$\Delta^{(n)}(k,r)$~:

$$\Delta^{(n)}(k, r) \leq \Delta_0(k) = \Delta (k, R) + {D \over k} \int_R^{\infty} |V(t)|dt
\quad , \eqno({\rm A}.24)$$

\noi valid for all $r > R$, including $r = \infty$, and all $n$. The sequence has therefore a
limit, and this limit provides the solution $\Delta (k, r)$ of (A.1) for all $r \geq R$,
including $r = \infty$, with $R < r_0(k)$, and $r_0(k)$ given by (A.16). Since we had also
proved the existence of the solution in $[0, r_0]$, we have therefore proved the existence of
the solution of (A.1) for all $r \geq 0$. Obviously, we have

$$\Delta (k, 0) = 0 \quad . \eqno({\rm A}.25)$$

\noi Therefore, we have, for $r > R$,

$$\Delta (k, r) \leq \Delta (k, R) + {D \over k} \int_R^{\infty} |V(t)| dt \quad . \eqno({\rm
A}.25)$$

\noi If we note also that $\Delta (k, r)$ is an increasing function of $r$, we secure the
existence of $\Delta (k) = \Delta (k, \infty )$, and from (A.22) and (A.24), we obviously
have, for the phase-shift

$$\lim_{k \to \, \infty} \delta (k) \leq \lim_{k \to \, \infty} \Delta (k) = \lim_{k \to
\, \infty} \Delta (k, \infty ) = 0 \quad . \eqno({\rm A}.26)$$ 

\noi This completes our proof for the general case where $rV(r) \in L^1(0, \infty )$.

   \newpage
\noindent
{\Large\bf Appendix B} \\

Here, we shall show that, for regular potentials satisfying (1), one
can make (3) more
precise if one knows the detail behaviour of the potential when $r
\to 0$. One has,
indeed, the following~: \\

\noi {\bf Theorem 1.}$^4$ Let us assume that $V$ is continuous and
bounded, away from
the origin, in $[0, R]$, $R < \infty$, and is such that

$$\lim_{r \to \ 0} r^{1 + \alpha} \ V(r) = V_0 \quad , \quad 0 <
\alpha < 1 \quad .
\eqno({\rm B}.1)$$

\noi We have then, as $k \to \infty$,

$$\delta (k) = {- V_0 \over \alpha} \cos \left ( {\pi \alpha \over 2}
\right ) \Gamma
(1 - \alpha ) {1 \over (2k)^{1 - \alpha}} + \cdots \quad , \eqno({\rm B}.2)$$

\noi and conversely~: (B.2) entails (B.1). \par

We shall give the proof for the $S$-wave, $\ell = 0$. The proof for
higher waves is
quite similar. Because of (14), we can just limit ourselves to (29)~:

$$\delta (k) \cong -  \int_0^R dt \sin 2 kt \left ( \int_t^R V(r) \
dr \right ) \quad
. \eqno({\rm B}.3)$$

\noi For the asymptotic behaviour of this integral (as $k \to
\infty$), we can now use
the following theorem of Titchmarsh$^{11}$~: \\

\noi {\bf Theorem 2.} Let $f(x)$ and $f'(x)$ be integrable over any
finite interval
not ending at $x = 0$~; let $x^{1 + \alpha} \ f'(x)$ be bounded for
all $x$, and let
$f(x) \sim x^{-\alpha}$ as $x \to 0$. Then, denoting by $F_0$ the limit of $x^{1+ \alpha}f'(x)$
as $x \to 0$, we have 

$$\int_0^{\infty} f(x) \sin k x \ dx = F_0 \ \Gamma (1 - \alpha ) \cos {\pi
\alpha \over 2}
\ {1 \over k^{1 - \alpha}} (1 + o(1)) \quad , \eqno({\rm B}.4)$$

\noi as $k \to \infty$. The converse theorem is also true if we deal
with finite
intervals in (B.4) since we can define the Fourier inverse transforms in a
straightforward manner. This is indeed the case in (B.3). \par

Our theorem 1 follows now immediately from the theorem of Titchmarsh applied to
(B.3). Moreover, it is a Tauberian kind theorem, i.e. its converse is
also true if we
remember that the Fourier transform of an $L^1$ function $f(x)$ is a continuous
function $F(k)$ of $k$. More refined theorems containing logarithmic
terms can also
be proved. Examples are treated in Calogero's book$^{5}$. For higher
waves, the proof
is similar by using (39.b) and (43) in (62).

\newpage

\noindent
{\Large\bf References}

\ben
   \item R. G. Newton, Scattering Theory of Waves and Particles, 2nd ed.
(Springer-Verlag, Berlin, 1982).
 
\item A. Galindo and P. Pascual, Quantum Mechanics, 2 volumes
(Springer-Verlag, Berlin, 1990).
 
\item K. Chadan and P. Sabatier, Inverse Problems in Quantum Scattering
Theory, 2nd ed. (Springer-Verlag, Berlin, New York 1989).

\item K. Chadan in K. Chadan, D. Colton, L. P\"{a}iv\"{a}rinta and W.
Rundell,  An Introduction to Inverse Scattering and Inverse Spectral Problems
(SIAM, Philadelphia 1997).

\item  F. Calogero, Variable Phase Approach to Potential Scattering
(Academic Press, New York 1967). See especially chapter 3, and
chapter 6, formulae
(3) and (8). A complete list of references is given in this book.

   \item M. Abramowitz and I. A. Stegun, Handbook of Mathematical Functions.

\item W. Frank and D. J. Land, Singular Potentials,
Rev. Mod. Phys. \underline{43}, 36-98 (1971).

    \item N. Fr\"{o}man and K. E. Thylwe, Setting the question of the high
energy behaviour of phase-shifts $\cdots$ ,
J. Math. Phys. \underline{20}, 1716-1719 (1979).
 
\item O. Brander, High-energy behaviour of phase-shifts for scattering
from singular potentials, J. Math. Phys. \underline{22}, 1229-1235
(1981).
 
\item K. Chadan, N. N. Khuri, A. Martin and T. T. Wu, Universality of
low-energy scattering in $2 + 1$ dimensions, Phys. Rev.
D\underline{58}, 025014-1-19
(1998).

\item E. C. Titchmarsh, Theory of Fourier integrals, 2nd ed. (Oxford University
Press, Oxford 1962), pp 172-173, theorems 126 and 127 combined together.

\item E. Hille, Lectures on Ordinary Differential Equations (Addison-Wesley,
Reading, MA, 1969).

\item E.-A. Coddington and N. Levinson, Theory of Ordinary
Differential Equations
(McGraw-Hill, New York, 1955).
\een

\end{document}